\documentclass[aps,twocolumn,showpacs,preprintnumbers,amsmath,amssymb]{revtex4-1}
\usepackage{graphicx}% Include figure files
\usepackage{dcolumn}% Align table columns on decimal point
\usepackage{bm}% bold math
\usepackage{color}

\begin{document}

\title{Quantum-disordered state of magnetic and electric dipoles\\
in a hydrogen-bonded Mott system}

\author{M.~Shimozawa$^{1,*,{\dag}}$}
\author{K.~Hashimoto$^{2,*,{\dag}}$}
\author{A.~Ueda$^1$}
\author{Y.~Suzuki$^1$}
\author{K.~Sugii$^1$}
\author{S.~Yamada$^1$}
\author{Y.~Imai$^1$}
\author{R.~Kobayashi$^2$}
\author{K.~Itoh$^2$}
\author{S.~Iguchi$^2$}
\author{M.~Naka$^3$}
\author{S.~Ishihara$^3$}
\author{H.~Mori$^1$}
\author{T.~Sasaki$^2$}
\author{M.~Yamashita$^1$}

\affiliation{
$^1$\mbox{The Institute for Solid State Physics, The University of Tokyo, Kashiwa, Chiba 277-8581, Japan}\\
$^2$\mbox{Institute for Materials Research, Tohoku University, Aoba-ku, Sendai 980-8577, Japan}\\
$^3$\mbox{Department of Physics, Tohoku University, Sendai 980-8578, Japan}\\
$^*$These authors contributed equally to this work.\\
$^{\dag}$\rm{To whom correspondence should be addressed.\\E-mail: shimo@issp.u-tokyo.ac.jp (M.S.); hashimoto@imr.tohoku.ac.jp (K.H.)}
}

%\date{\today}

%\pacs{}

\maketitle

%%%%%%%%%  Abstract  %%%%%%%%%
{\bf Strongly enhanced quantum fluctuations %The dominance of quantum fluctuations over classical long-range order 
often lead to a rich variety of quantum-disordered states. A representative case is liquid helium, in which zero-point vibrations of the helium atoms prevent its solidification at low temperatures. A similar behaviour is found for the internal degrees of freedom in electrons. Among the most prominent is a quantum spin liquid (QSL)\cite{Balents10}, in which localized spins are highly correlated but fluctuate even at absolute zero. Recently, a coupling of spins with other degrees of freedom has been proposed as an innovative approach to generate even more fascinating QSLs such as orbital--spin liquids\cite{Feiner97}. However, such ideas are limited to the internal degrees of freedom in electrons. Here, we demonstrate that a coupling of localized spins %$\pi$-electrons 
with the zero-point motion of hydrogen atoms (proton fluctuations) in a hydrogen-bonded organic Mott insulator\cite{Isono13,Isono14,Ueda14,Tsumuraya15} %a coupling between the zero-point motion of hydrogen atoms (proton fluctuations) and $\pi$-electrons %on a two-dimensional spin-1/2 Heisenberg triangular lattice 
provides a new class of QSLs. We find that a divergent dielectric behaviour towards a hydrogen-bond order is suppressed by the quantum proton fluctuations, resulting in a quantum paraelectric (QPE) state\cite{Muller79,Shen15,Barrett52}. Furthermore, our thermal-transport measurements reveal that a QSL state with gapless spin excitations rapidly emerges upon entering the QPE state. These findings indicate that the quantum proton fluctuations give rise to a novel QSL --- a quantum-disordered state of magnetic and electric dipoles --- through the coupling between the electron and proton degrees of freedom. %Utilizing such a coupling between the internal degrees of electrons and atoms may advance our explorations of novel quantum phenomena.
}

%%%%%%%%%  Introduction  %%%%%%%%%

%\UTF{2460}三角格子ではQSLを実現するのに量子揺らぎが不十分。スピンが電荷、軌道秩序などとカップルすることが量子揺らぎを強める上でキーポイントであることを示す。
The nature of QSLs has been well established in one-dimensional (1D) spin systems. However, it still remains elusive how QSLs emerge in dimensions greater than one. The celebrated resonating-valence-bond theory on a 2D triangular lattice \cite{Anderson73} puts forward the possibility that geometrical frustration plays an important role for stabilizing QSLs. In fact, a few candidate materials hosting QSLs have now been reported in materials with 2D triangular lattices \cite{Isono14,Shimizu03,Itou08,Shimizu16,Shen16,Paddison16}. Nevertheless, according to subsequent theoretical studies \cite{Bernu92,Capriotti99}, the effect of geometrical frustration in the triangular lattice is insufficient to stabilize QSLs, leading to %intensive efforts to search for other 
a number of mechanisms that may stabilize the QSL states found in the candidate materials. 
%\textcolor{red}{One of the most promising ideas is that novel QSLs are stabilized by utilizing a coupling of spins with charges and orbitals; the former has been discussed near a Mott-insulator-to-metal transition where the charge degrees of freedom begin to delocalize \cite{Morita02,Yoshioka09,Misguich99,Motrunich05,Mross11,Hotta10,Naka16}, and the latter has been considered in the framework of a dynamic Jahn--Teller effect \cite{Feiner97,Nakatsuji12}.}
One of the most promising approaches is to utilize a coupling of spins with %other degrees of freedom, such as 
charges and orbitals; the former has been discussed near a Mott-insulator-to-metal transition where the charge degrees of freedom begin to delocalize \cite{Morita02,Yoshioka09,Misguich99,Motrunich05,Mross11,Hotta10,Naka16} and the latter has been considered in the framework of a spin--orbital coupling %dynamic Jahn--Teller effect 
\cite{Feiner97}. %Nevertheless, given the limited amount of the experimental studies, the mechanisms stabilizing QSLs have remained unknown. 
Such strategies, however, have been limited to the utilization of internal degrees of freedom of electrons.
Here, we demonstrate %\textcolor{red}{an unprecedented} 
a novel approach for stabilizing a QSL realized in a hydrogen-bonded organic Mott insulator, in which a coupling between $\pi$-electrons and hydrogen atoms provides a novel quantum-disordered state of magnetic and electric dipoles.

%%%%%%%%%%%%%%%%% Figure 1 %%%%%%%%%%%%%%%%%%
\begin{figure*}[t]
\begin{center}
\includegraphics[width=0.8\linewidth]{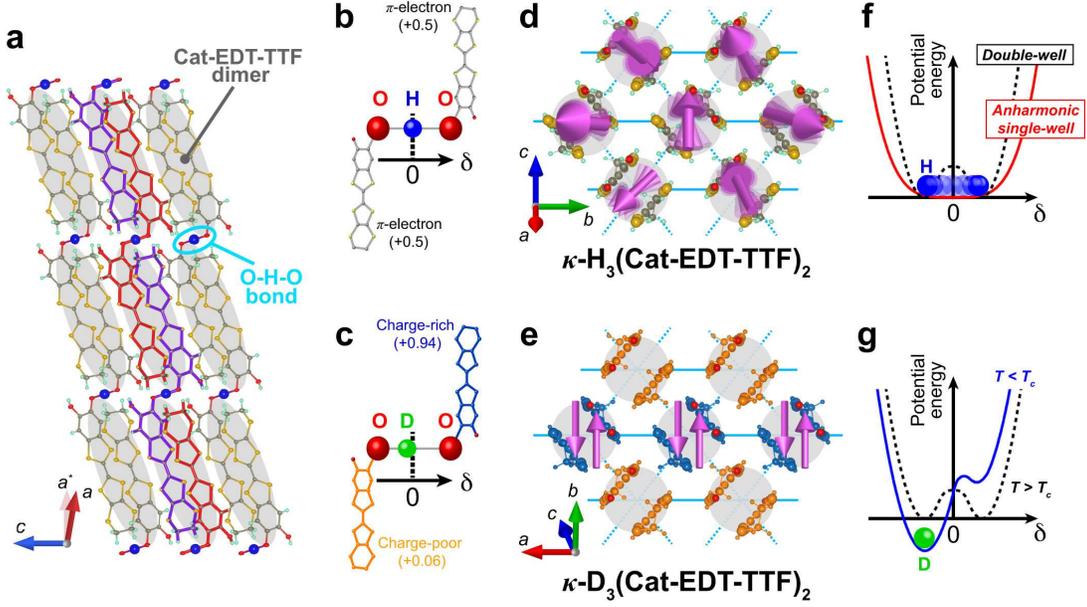}
\caption{Crystal structure and spin, charge and hydrogen/deuterium atom configurations in H-Cat and D-Cat.
{\bf a, b,} Crystal structure of H-Cat viewed along the $b$ axis ({\bf a}) and hydrogen-bonded molecular unit H$_3$(Cat-EDT-TTF)$_2$ (supramolecule) in H-Cat ({\bf b}). The supramolecules are formed by two [H(Cat-EDT-TTF)]$^{0.5+}$ molecules connected by the hydrogen bond. The supramolecules are stacked along the ($a+c$) direction, as shown in {\bf a}. For clarity, a part of the stacking columns is colored red and purple in {\bf a}. In the $b$-$c$ plane, two face-to-face [H(Cat-EDT-TTF)]$^{0.5+}$ molecules form a strongly dimerized unit (gray ellipsoid), which generates the 2D $\pi$-electron layers connected by the O-H-O hydrogen bonds (light blue circle). 
{\bf c,} Supramolecular unit in D$_3$(Cat-EDT-TTF)$_2$. Note that $\delta$ in {\bf b} ({\bf c}) denotes the displacement of the hydrogen (deuterium) atom from the center of the O-H-O (O-D-O) bond.
{\bf d,} Spin and charge structures of the 2D $\pi$-electron layer in H-Cat. The $\pi$-dimers form a slightly anisotropic triangular lattice with $S = 1/2$ spins \cite{Isono13} (magenta arrows). 
{\bf e,} Spin and charge structures of a 2D $\pi$-electron layer in D-Cat below the phase transition temperature $T_c$ of 185 K. Charge disproportionation associated with deuterium localization leads to a non-magnetic ground state below $T_c$. The blue- and orange-colored dimers indicate the charge-rich ($+0.94$) and charge-poor ($+0.06$) sites, respectively \cite{Ueda14}. 
{\bf f, g,} Schematics of the potential energy curves of the hydrogen atoms in H-Cat ({\bf f}) and the deuterium atoms in D-Cat ({\bf g}). In H-Cat, the potential energy curve is suggested to change from a double-well structure (dashed line) to a very shallow and anharmonic single-well structure (red solid line) owing to the many-body effect arising from the network of hydrogen bonds and $\pi$-electrons \cite{Tsumuraya15,Yamamoto16}. In sharp contrast, the energy curve in D-Cat retains a double-well structure above $T_c = 185$ K (dashed line), leading to the deuterium localization at $T_c$ (blue solid line).}
\label{Fig1}
\end{center}
\vspace{-5mm}
\end{figure*}
%%%%%%%%%%%%%%%%% Figure 1 %%%%%%%%%%%%%%%%%%

%\UTF{2461}Cat-EDT-TTFに関する紹介１
$\kappa$-H$_3$(Cat-EDT-TTF)$_2$ (hereafter abbreviated as H-Cat) is a hydrogen-bonded organic Mott insulator, where Cat-EDT-TTF is catechol-fused ethylenedithiotetrathiafulvalene (refs\,\onlinecite{Isono13,Isono14,Ueda14,Tsumuraya15}). H-Cat forms a 2D spin-1/2 Heisenberg triangular lattice of Cat-EDT-TTF dimers \cite{Isono13} (Figs\,\ref{Fig1}\,a and d). Despite the antiferromagnetic interaction energy $J/k_{\rm B}$ of $\sim80$ K, no magnetic order has been observed down to 50 mK; this indicates the realization of a QSL state \cite{Isono14}. A distinct feature of H-Cat is that the 2D $\pi$-electron layers are connected by hydrogen bonds \cite{Isono13,Ueda14} (Figs\,\ref{Fig1}\,a and b), which is in marked contrast to other 2D organic QSL materials such as $\kappa$-(BEDT-TTF)$_2$Cu$_2$(CN)$_3$ (ref.\,\onlinecite{Shimizu03}) and EtMe$_3$Sb[Pd(dmit)$_2$]$_2$ (ref.\,\onlinecite{Itou08}) where the 2D spin systems are separated by non-magnetic insulating layers. This structural feature of H-Cat is highlighted by deuteration of the hydrogen bonds \cite{Ueda14}; specifically, in the deuterated analogue of H-Cat, $\kappa$-D$_3$(Cat-EDT-TTF)$_2$ (denoted as D-Cat), deuterium localization occurs at $T_c = 185$ K, accompanied by charge disproportionation within the Cat-EDT-TTF layers, resulting in a non-magnetic ground state (Figs\,\ref{Fig1}c and e). This demonstrates that the hydrogen bonds in this system strongly couple with the charge and spin degrees of freedom of the $\pi$-electrons in the Cat-EDT-TTF dimers.

%\UTF{2462}Cat-EDT-TTFに関する紹介２
In contrast to D-Cat, the hydrogen atoms in H-Cat do not localize down to low temperatures \cite{Ueda14}. This is inconsistent with the fact \cite{Yamamoto16} that the potential energy curve of the hydrogen bonds calculated for an isolated supramolecule has a double minimum potential with a large energy barrier of $\sim 800$ K (see Figs\,\ref{Fig1}b and f), which should localize the hydrogen atoms in H-Cat at low temperatures. Recent theoretical calculations \cite{Tsumuraya15,Yamamoto16} have pointed out that the potential energy curve has a single-well structure and its bottoms become very shallow and anharmonic (see Fig.\,\ref{Fig1}f) owing to a many-body effect arising from the network of hydrogen bonds and $\pi$-electrons. In this case, the zero-point motions of the hydrogen atoms (termed ``proton fluctuations") can be strongly enhanced by the anharmonic potential curve. In contrast to D-Cat, the enhanced proton fluctuations may delocalize the hydrogen atoms down to absolute zero, providing a playground for realizing the QSL state through the strong coupling between the hydrogen bonds and the $\pi$-electrons. However, it has been elusive whether such strong quantum proton fluctuations are indeed present in H-Cat, and if so, how the quantum fluctuations affect the QSL state.

%%%%%%%%%  Experimental  %%%%%%%%%

%\UTF{2463}測定手法について（誘電率と熱伝導率）
To answer these questions, we demonstrate a combination of dielectric permittivity and thermal conductivity measurements. These methods are particularly suitable because the dielectric permittivity is sensitive to local electric-dipole moments arising from the hydrogen-bond dynamics \cite{Horiuchi03}, whereas the thermal conductivity is a powerful probe to detect itinerant low-lying energy excitations associated with the nature of QSL states \cite{Yamashita09,Yamashita10}.

%%%%%%%%%%%%%%%%% Figure 2 %%%%%%%%%%%%%%%%%%
\begin{figure*}[t]
\begin{center}
\includegraphics[width=0.65\linewidth]{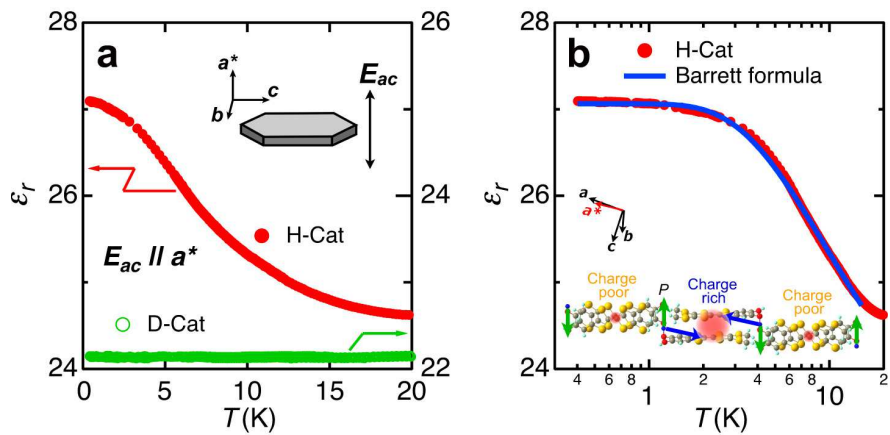}
\caption{Dielectric permittivity of H-Cat and D-Cat.
{\bf a,} Temperature dependence of the dielectric constant $\epsilon_r(T)={\epsilon}/{\epsilon_0}$ in H-Cat (red, left axis) and D-Cat (green, right axis), where $\epsilon$ is the dielectric permittivity and $\epsilon_0$ is the vacuum permittivity. The inset illustrates our measurement configuration. The dielectric permittivity was measured by applying an a.c. electric field ($E_{ac}$) of 1 MHz along the $a^*$ axis. 
{\bf b,} The same data for H-Cat plotted on a logarithmic temperature scale. The solid line is a fit to the Barrett formula (see the main text), demonstrating the emergence of a QPE state in H-Cat. The inset shows the electric dipole moments induced within the deuterium bonds and the Cat-EDT-TTF dimers in D-Cat. Because of deuterium localization, the adjacent deuterium bonds possess the local electric dipoles oriented in antiparallel directions (green arrows) below $T_c$. Concomitantly, charge disproportionation is induced within the Cat-EDT-TTF dimers (Fig.\,1c), giving rise to the local electric dipoles (blue arrows). Thus, the antiferroelectric dipole--dipole interactions are induced by the hydrogen/deuterium-bond dynamics.}
\label{Fig2}
\end{center}
\vspace{-5mm}
\end{figure*}
%%%%%%%%%%%%%%%%% Figure 2 %%%%%%%%%%%%%%%%%%

%%%%%%%%%  Main topic  %%%%%%%%%

%\UTF{2464}誘電率測定の結果
Figure\,\ref{Fig2} shows the temperature dependence of the dielectric constant $\epsilon_r(T)$ for H-Cat and D-Cat. In H-Cat, $\epsilon_r(T)$ steeply increases with decreasing temperature and then saturates below $\sim 2$ K. In sharp contrast, $\epsilon_r(T)$ of D-Cat is temperature-independent owing to deuterium localization (Fig.\,\ref{Fig2}a). The temperature dependence of $\epsilon_r$ for H-Cat is a typical dielectric behaviour observed in quantum paraelectric (QPE) materials such as SrTiO$_3$ (ref.\,\onlinecite{Muller79}), in which long-range electric order is suppressed by strong quantum fluctuations. In the QPE state, $\epsilon_r(T)$ is described by the so-called Barrett formula \cite{Barrett52}:
\begin{equation}\label{Eq:Barrett}
\epsilon_r(T) = A + \frac{C}{\frac{T_1}{2}\coth(\frac{T_1}{2T})-T_0},
\end{equation}
where $A$ is a constant offset, $C$ the Curie constant, $T_0$ the Curie-Weiss temperature in the classical limit (proportional to the effective dipole--dipole coupling strength) and $T_1$ the characteristic crossover temperature from the classical to the quantum regime. As shown in the solid line in Fig.\,\ref{Fig2}b, $\epsilon_r(T)$ of H-Cat is well fitted by the Barrett formula with $T_0=-6.4$ K and $T_1=7.7$ K; this confirms that strong quantum fluctuations suppress long-range electric order. The obtained negative value of $T_0$ immediately indicates the presence of an antiferroelectric (AFE) dipole--dipole interactions in H-Cat, which is consistent with the static AFE dipole moments arising from deuterium localization in D-Cat (see the inset of Fig.\,\ref{Fig2}b). Therefore, the observed quantum paraelectricity in H-Cat clearly shows that strong quantum fluctuations that suppress the hydrogen-bond order as observed in D-Cat arise from the potential energy curve of H-Cat, consequently leading to the persistence of enhanced proton fluctuations down to low temperatures. The presence of strong quantum fluctuations is consistent with the recent theoretical calculations that highlight the importance of strong many-body effects imposed by the proton--$\pi$-electron network on the potential energy curve of H-Cat \cite{Tsumuraya15,Yamamoto16}.

%%%%%%%%%%%%%%%%% Figure 3 %%%%%%%%%%%%%%%%%%
\begin{figure*}[t]
\begin{center}
\includegraphics[width=0.65\linewidth]{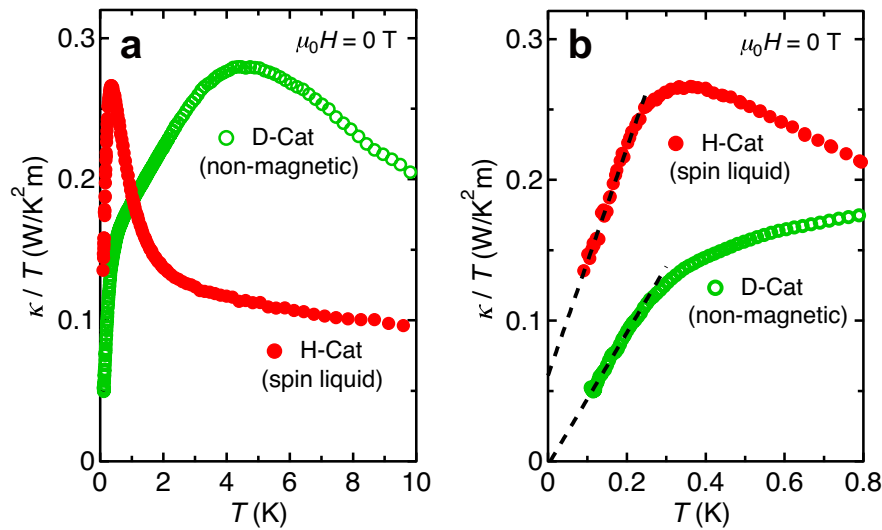}
\caption{Thermal conductivity of H-Cat and D-Cat.
{\bf a,} Temperature dependence of $\kappa/T$ of H-Cat (filled red symbols) and D-Cat (open green symbols) below 10 K in the zero-field. 
{\bf b,} Low-temperature thermal conductivity for H-Cat and D-Cat below 0.8 K. The extrapolated data (dashed lines) assume that $\kappa_{\rm ph}\propto T^2$ (for details, see Supplementary Information Sec.\,II). A clear residual $\kappa/T$ in the zero-temperature limit is resolved for H-Cat.}
\label{Fig3}
\end{center}
\vspace{-5mm}
\end{figure*}
%%%%%%%%%%%%%%%%% Figure 3 %%%%%%%%%%%%%%%%%%

%\UTF{2465}熱伝導率測定の結果
Using thermal conductivity measurements, we next examine how the proton dynamics in the QPE state affects the nature of the QSL state in H-Cat. Figure\,\ref{Fig3} shows the temperature dependence of the thermal conductivity of H-Cat ($\kappa^{\rm H}$) and D-Cat ($\kappa^{\rm D}$). The heat in H-Cat is carried by the phonons ($\kappa_{\rm ph}^{\rm H}$) and the spin excitations ($\kappa_{\rm sp}^{\rm H}$), whereas in non-magnetic D-Cat, it is transported only by phonons ($\kappa_{\rm ph}^{\rm D}$). Assuming that H-Cat and D-Cat share almost identical phonon thermal conductivity ($\kappa_{\rm ph}^{\rm H}\sim\kappa_{\rm ph}^{\rm D}$), the relation $\kappa^{\rm H}=\kappa_{\rm ph}^{\rm H}+\kappa_{\rm sp}^{\rm H}\geq\kappa_{\rm ph}^{\rm D}=\kappa^{\rm D}$ holds. Unexpectedly, however, we find that $\kappa^{\rm H}<\kappa^{\rm D}$ above 2 K (see Fig.\,\ref{Fig3}a), indicating that $\kappa_{\rm ph}^{\rm H}$ is much more suppressed than $\kappa_{\rm ph}^{\rm D}$.

\begin{figure*}[t]
\begin{center}
\includegraphics[width=0.8\linewidth]{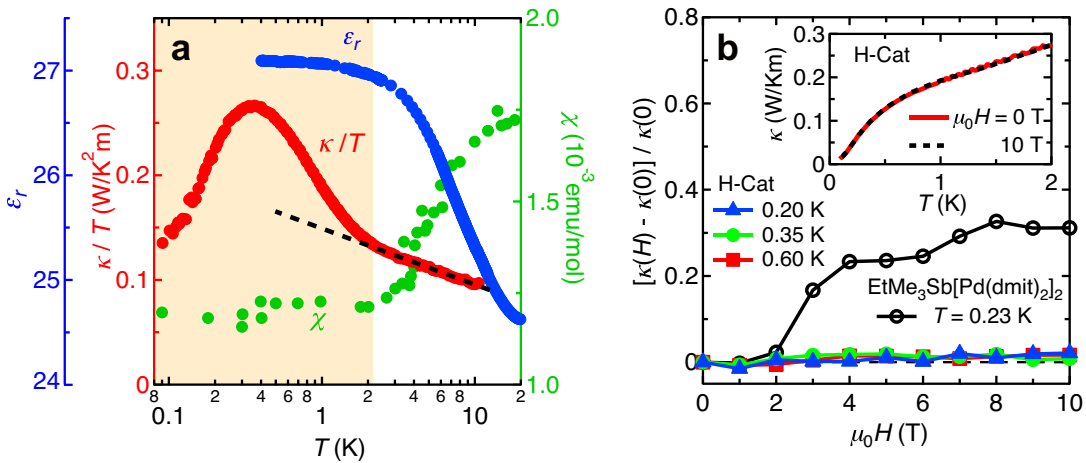}
\caption{QPE and QSL states of H-Cat.
{\bf a,} A combination of the temperature dependences of the dielectric constant $\epsilon_r$ (blue, left axis), the thermal conductivity divided by temperature $\kappa/T$ (red, left axis) and the magnetic susceptibility $\chi$ (green, right axis) for H-Cat. The values of $\chi$ are taken from Ref.\,\onlinecite{Isono14}. The dashed line is an eye guide. 
{\bf b,} Magnetic field dependence of the thermal conductivity $\kappa(H)$ normalized by the zero field value $[\kappa(H)-\kappa(0)]/\kappa(0)$ for H-Cat (filled symbols) and EtMe$_3$Sb[Pd(dmit)$_2$]$_2$ (open circles) \cite{Yamashita10}. The inset shows the temperature dependences of $\kappa(H)$ for H-Cat at 0 T and 10 T.}
\label{Fig4}
\end{center}
\vspace{-5mm}
\end{figure*}

%\UTF{2466}H体で熱伝導率が抑制される原因
To investigate the origin of this suppression, we employ the Callaway model \cite{Berman76}, which describes the heat transport of acoustic phonons. Above 2 K, $\kappa^{\rm H}$ is reproduced by the model including a single resonance scattering mode with a resonance energy of $\hbar\omega_0/k_{\rm B}\sim$ 5--10 K in addition to standard scattering processes (see Supplementary Information Sec.\,I). This energy scale is close to the proton fluctuations ($T_1=7.7$ K), indicating that resonance scattering arises between the acoustic phonons and the optical mode from the hydrogen bonds. Here, it should be noted that we can rule out the possibility of a spin--phonon scattering for the suppression of $\kappa_{\rm ph}^{\rm H}$, because the spin--orbit coupling of H-Cat is very weak, as confirmed by the small field dependence of $\kappa^{\rm H}$ (see the inset of Fig.\,\ref{Fig4}b). Thus, it appears that the thermal fluctuations of the hydrogen bonds strongly suppress $\kappa^{\rm H}$ above 2 K.

%\UTF{2467}低エネルギー励起（QSL）
%We now focus on the low-lying excitation spectrum characterizing the QSL state, which can be inferred from the thermal conductivity in the lower temperature regime\cite{Yamashita09,Yamashita10}. 
Below 2 K, $\kappa^{\rm H}$ rapidly increases and eventually exceeds $\kappa^{\rm D}$ (Fig.\,\ref{Fig3}a). This rapid increase of $\kappa^{\rm H}$ may come from an increase in $\kappa_{\rm sp}^{\rm H}$. We now investigate the behaviour of $\kappa_{\rm sp}^{\rm H}$ at lower temperatures, which provides essential information on the low-lying excitation spectrum characterizing the QSL state \cite{Yamashita09,Yamashita10}. As shown in the dotted lines in Fig.\,\ref{Fig3}b, both $\kappa_{\rm ph}^{\rm H}$ and $\kappa_{\rm ph}^{\rm D}$ exhibit a $T^2$-dependence rather than the conventional $T^3$-dependence. This is originated from the influence of high-quality crystals with specular surfaces (see Supplementary Information Sec.\,II). The zero-temperature extrapolation of $\kappa^{\rm H}/T$ shows a finite residual (Fig.\,\ref{Fig3}b), thereby demonstrating a gapless spin excitation with high mobility (the mean free path of the gapless spin excitations $l_{\rm sp}$ is estimated to be $\sim 120$ nm; See Supplementary Information Sec.\,III). This result is consistent with recent magnetic torque measurements \cite{Isono14} of H-Cat.

%\UTF{2468}QSLがどのようにして実現するか？これまでの一般論
A key question raised here is how the gapless QSL state is stabilized in H-Cat. In organic QSL candidates with a triangular lattice, charge fluctuations near a Mott transition \cite{Morita02,Yoshioka09,Misguich99,Motrunich05,Mross11,Hotta10,Naka16} have been pointed out to play an important role for stabilizing the QSLs. However, H-Cat is located deeper inside the Mott-insulating phase as compared to the other organic QSL candidates, $\kappa$-(BEDT-TTF)$_2$Cu$_2$(CN)$_3$ (ref.\,\onlinecite{Shimizu03}) and EtMe$_3$Sb[Pd(dmit)$_2$]$_2$ (refs\,\onlinecite{Itou08,Itou10}). The distance from a Mott transition is inferred from the ratio of the on-site Coulomb repulsion $U$ to the transfer integral $t$, which is given by $U/t\sim t/J$. Whereas the transfer integrals are comparable among the three materials, $J$ for H-Cat is $\sim 1/3$ compared to that for the other two (see Supplementary Information Sec.\,IV); this means that $U/t$ is significantly large in H-Cat. Indeed, H-Cat sustains an insulating behaviour even at 1.6 GPa (ref.\,\onlinecite{Isono13}), whereas the other two compounds become metallic at 0.4--0.6 GPa (ref.\,\onlinecite{Furukawa15}); this result also supports that H-Cat is far from the Mott transition. Therefore, the QSL in H-Cat should be stabilized using a completely different mechanism.

%%%%%%%%%%%%%%%%% Figure 4 %%%%%%%%%%%%%%%%%%
%%%%%%%%%%%%%%%%% Figure 4 %%%%%%%%%%%%%%%%%%

%\UTF{2469}実験+理論：プロトンの揺らぎとスピンの相関について（QSLとQPE）
Figure\,\ref{Fig4}a shows the temperature dependence of the dielectric constant $\epsilon_r$, the thermal conductivity divided by temperature $\kappa/T$ and the magnetic susceptibility $\chi$ (ref.\,\onlinecite{Isono14}) for H-Cat. Below 2 K, the thermal conductivity increases upon entering the QPE state, where $\epsilon_r$ saturates. This characteristic temperature coincides with the temperature at which the susceptibility becomes constant; this occurs when the spin correlation develops in the QSL state \cite{Isono14}. This coincidence of the QPE and QSL states is surprising and strongly suggests that the development of the quantum proton fluctuations triggers the emergence of the QSL. We now theoretically analyze the effects of proton dynamics on the QSL state. In H-Cat, the charge degrees of freedom of the hydrogen bonds and the $\pi$-electrons are strongly coupled because of the charge neutrality within the H$_3$(Cat-EDT-TTF)$_2$ supramolecule (see Figs\,\ref{Fig1}b and c). A minimal and realistic model that describes the coupling between $\pi$-electrons and hydrogen bonds is the extended Hubbard model coupled with the proton degree of freedom; this model captures the essence of this system, namely that the proton and charge degrees of freedom are strongly entangled in H-Cat (see Supplementary Information Sec.\,V). According to this model, charge fluctuations between the two molecules inside a dimer are governed by proton fluctuations ($\sim$ 1 meV), which are comparable to the magnetic exchange interaction $J$ in the $\pi$-electron spin system. Such low-energy charge fluctuations lead to a dynamical modulation of $J$ as well as a reduction of the on-site Coulomb repulsion $U$ due to the bi-polaron effect \cite{Alexandrov96}, both of which may destabilize the magnetic long-range order, that is, induce a QSL state. 

%A minimal model that describes the coupling between the $\pi$-electrons and proton degrees of freedom is the 1D extended Hubbard model (see Supplementary Information Sec.\,V). This model captures an important aspect that the proton and charge (hole) degrees of freedom strongly entangle each other in H-Cat, which may induce intra/inter-dimer charge fluctuations within the 2D $\pi$-electrons and stabilize the QSL state. 

%\UTF{246A}プロトンと電子が相互作用することでこれまでとは違うエキゾチックな量子相(an exotic quantum phase of matter)が現れた。磁場依存性から評価。

Finally, we discuss the magnetic field dependence of the thermal conductivity $\kappa(H)$ in the novel QSL state of H-Cat. As shown in Fig.\,\ref{Fig4}b, the field dependence of $\kappa(H)$ is much weaker in H-Cat than in $\kappa$-(BEDT-TTF)$_2$Cu$_2$(CN)$_3$ (ref.\,\onlinecite{Yamashita09}) and EtMe$_3$Sb[Pd(dmit)$_2$]$_2$ (ref.\,\onlinecite{Yamashita10}). The magnetic field dependence of the thermal conductivity in $\kappa$-(BEDT-TTF)$_2$Cu$_2$(CN)$_3$ and EtMe$_3$Sb[Pd(dmit)$_2$]$_2$ has been discussed in terms of an inhomogeneous QSL \cite{Pratt11} and a dichotomy of gapless and gapped excitations \cite{Yamashita10}, respectively. Conversely, the observed small field dependence of the thermal conductivity in H-Cat suggests that a more globally homogeneous QSL with gapless excitations is realized in this system. This difference may %come from the different mechanisms stabilizing the QSL state, as discussed above. These results 
highlight the unique feature of H-Cat, namely, that the QSL is induced by quantum proton fluctuations rather than charge fluctuations near a Mott transition, as discussed above. Thus, our findings suggest that a novel quantum entangled state, that is, a quantum-disordered state of magnetic and electric dipoles emerges in H-Cat from cooperation between the electron and proton degrees of freedom. Utilizing such a strong coupling between multiple degrees of freedom will advance our explorations of quantum phenomena such as orbital--spin liquids \cite{Feiner97,Nakatsuji12} and electric-dipole liquids \cite{Shen15,Hotta10}.

%%%%%%%%%%%%%%	METHODS
\section*{Methods}

{\bf Sample preparation.}
Single crystals of $\kappa$-H$_3$(Cat-EDT-TTF)$_2$ and $\kappa$-D$_3$(Cat-EDT-TTF)$_2$ were prepared by the electrochemical oxidation method, as described in refs \onlinecite{Isono13} and \onlinecite{Ueda14}. A typical sample size for both compounds is $\sim 0.03$ mm $\times \, 0.12$ mm $\times \, 1.0$ mm.

{\bf Dielectric measurements.}
The dielectric permittivity measurements were carried out down to 0.4 K in a $^3$He cryostat using an LCR meter (Agilent 4980A) operated at 1 MHz along the $a^*$ direction, which is perpendicular to the $b$-$c$ plane. The experiment was limited to the $a^*$-axis direction by the plate-like shape of the sample. %The magnetic field parallel to the $\vect{a^*}$ direction was applied up to 15 T.

{\bf Thermal conductivity measurements.}
The thermal-transport measurements were performed by a standard steady-state heat-flow technique in the temperature range from 0.1 to 10 K using a dilution refrigerator. The heat current was applied along the $c$ axis. The magnetic field was applied perpendicularly to the $b$-$c$ plane up to 10 T. We attached two precisely calibrated RuO$_2$ thermometers in the magnetic field and one heater on the sample. 
%Since all samples studied here show similar characteristics for both $\kappa$-H$_3$(Cat-EDT-TTF)$_2$ and $\kappa$-D$_3$(Cat-EDT-TTF)$_2$, we can exclude the possibility of sample inhomogeneity and the influence of the contact-resistance.

%%%%%%%%%Acknowledgments
\section*{Acknowledgements}

We thank J. M{\"u}ller, M. Oshikawa, H. Seo, T. Shibauchi, M. Tachikawa, Y. Tada, T. Tsumuraya, H. Watanabe and K. Yamamoto for fruitful discussions. We also thank K. Torizuka and Y. Uwatoko for providing technical assistance. This work was supported by Grants-in-Aid for Scientific Research (Grants No.\,24340074, No.\,26287070, No.\,26610096, No.\,15H00984, No.\,15H00988, No.\,15H02100, No.\,15K13511, No.\,15K17691, No.\,16H00954, No.\,16H04010, No.\,16K05744 and No.\,16K17731) from MEXT and JSPS, by a Grant-in-Aid for Scientific Research on Innovative Areas ``$\pi$-Figuration'' (No. 26102001), by the Canon Foundation and by Toray Science Foundation.

%%%%%%%%%Author contributions
\section*{Author contributions}

M.S., K.H. and M.Y. conceived the project. M.S., Y.S., K.S., S.Y., Y.I. and M.Y. performed the thermal conductivity measurements. K.H., R.K., K.I., S.Iguchi and T.S. performed the dielectric permittivity measurements. A.U. and H.M. carried out sample preparation. M.N. and S.Ishihara gave theoretical advice. M.S., K.H. and M.Y. analyzed the data and wrote the manuscript with inputs from M.N. and S.Ishihara. All authors discussed the experimental results.

%%%%%%%%%References

\end{document}

%\preprint{submitted}

\begin{center}
{\large \bf Supplementary Information:\\
Quantum-disordered state of magnetic and electric dipoles in a hydrogen-bonded Mott system}
\end{center}

%%%%%%%%%%%%%%%%%%%%%%%%%%%%%%%%%%%%%%%%%%%%%%%%%%%%%%%%%%%%%%%%%%
%%%%%%%%%%%%%%%%%%%%%%%%%%%%%%%%%%%%%%%%%%%%%%%%%%%%%%%%%%%%%%%%%%
\section{Resonance scattering arising from hydrogen-bond dynamics}
%%%%%%%%%%%%%%%%%%%%%%%%%%%%%%%%%%%%%%%%%%%%%%%%%%%%%%%%%%%%%%%%%%
%%%%%%%%%%%%%%%%%%%%%%%%%%%%%%%%%%%%%%%%%%%%%%%%%%%%%%%%%%%%%%%%%%

To investigate how the phonon thermal conductivity is suppressed in H-Cat, we applied the Callaway model that describes a phonon thermal conductivity $\kappa_{\rm ph}$ \cite{Berman1976}: 
\begin{align}
\kappa_{\rm {ph}} = \frac{k_{\rm B}}{2\pi^2 v_{\rm ph}}\Bigl(\frac{k_{\rm B}}{\hbar}\Bigr)^3 T^3 \int_{0}^{\mathit{\Theta}_{\rm D}/T} \frac{x^4 e^x}{(e^2-1)^2}\tau(\omega,T)dx, \tag{S1}
\label{EqS1}
\end{align}
where $v_{\rm ph}$ is the average sound velocity of the phonons, $\mathit{\Theta}_{\rm D}$ is the Debye temperature, $\tau$ is the total phonon relaxation rate and $\omega$ is the frequency of phonons; moreover, $\hbar$ and $k_{\rm B}$ are 
the reduced Planck and Bolzmann constants, respectively, and $x=\frac{\hbar \omega}{k_{\rm B} T}$. By fitting the Debye model to the specific-heat data, the values of $v_{\rm ph}$ and $\mathit{\Theta}_{\rm D}$ were estimated as 1.7 km/s and 220 K, respectively. %\cite{S.Yamashita2016}. 
$\tau(\omega,T)$ was approximated as \cite{Sologubenko2001,Gofryk2014}:
\begin{align}
1/\tau(\omega,T) & = \frac{v_{\rm ph}}{L} + A\omega +D \omega^4 + B\omega^2 T \exp{\Bigl(-\frac{\Theta_{\rm D}}{f T}\Bigr)} \notag \\
& + g \Bigl(\frac{ \omega^4}{\omega^2-\omega_0^2}\Bigr)^2 \Bigl[1-h\tanh^2{\Bigl(\frac{\hbar\omega_0}{2k_{\rm B} T}\Bigr)}\Bigr], \tag{S2}
\label{EqS2}
\end{align}
where the first three terms represent the phonon scatterings at the sample boundaries, dislocations and point defects, respectively, and the last two terms stand for the scatterings caused by phonon--phonon (Umklapp) processes and a resonance mode, respectively. Here $L$, $A$, $D$, $B$, $f$, $g$ and $h$ are constant coefficients and $\omega_0$ is the resonance frequency. Using this model, we evaluated the phonon thermal conductivity of H-Cat above 2 K, where the spin contribution $\kappa_{\rm sp}$ is small. For comparison, we also analyzed the thermal conductivity of D-Cat in the same temperature range.

As shown in Fig.\,S1, $\kappa(T)$ for D-Cat is fitted by the above $\tau(\omega,T)$ without a resonance mode. In H-Cat, $\kappa(T)$ is well reproduced only by adding a single resonance scattering process with an energy of $\hbar \omega_0/k_{\rm B}\sim$ 8 K to the fitting result of D-Cat. The fitting parameters (summarized in Table\,S\ref{TableS1}) are in good agreement with those of other quantum spin systems \cite{Sologubenko2001}. This result indicates that the resonance mode strongly suppresses the phonon thermal conductivity of H-Cat. As H-Cat possesses both proton and spin degrees of freedom, this extra resonance scattering comes from either proton--phonon scattering or spin--phonon scattering. The resonance scattering energy almost coincides with the proton fluctuations ($T_1 = 7.7$ K), which suggests that the resonance scattering arises from the optical mode associated with the hydrogen bonds. Here, we can safely exclude spin--phonon scattering as the origin of the resonance scattering because of the following reasons. The spin--phonon scattering should be strongly quenched by Zeeman splitting under a magnetic field, which changes the field dependence of the thermal conductivity $\kappa(H)$. However, $\kappa(H)$ does not strongly depend on the magnetic field up to 10 T (see the inset of Fig.\,4b). Moreover, the $g$-value only slightly deviates from 2 and thereby supports that the spin--orbit coupling causing the spin--phonon scattering is very weak \cite{IsonoPRL2014}. Thus, we conclude that the strong suppression of $\kappa_{\rm ph}$ for H-Cat results from resonance scattering, which itself arises from the hydrogen-bond dynamics.

%%%%%%%%%%%%%%%%%%%%%%FIG S1%%%%%%%%%%%%%%%
\begin{figure}[t]
\begin{center}
\includegraphics[width=0.9\linewidth]{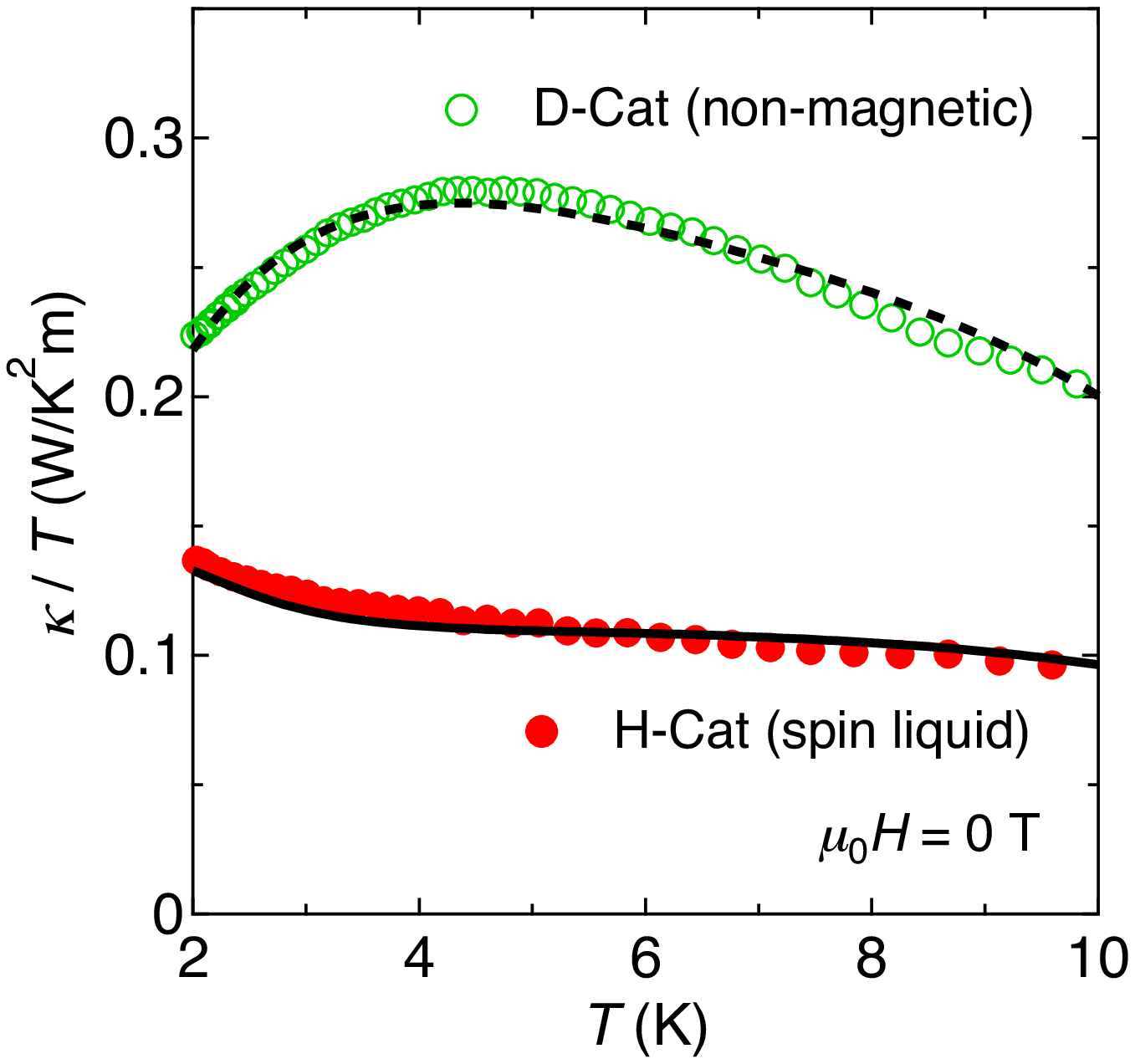}
\end{center}
\vspace{-5mm}
\caption{Temperature dependence of $\kappa/T$ for H-Cat (filled red symbols) and D-Cat (open green symbols) over the temperature range 2--10 K in 
the zero-field. Solid and dashed lines are the Callaway model fits to H-Cat and D-Cat data, respectively.}
\label{FigS1}
\vspace{-3mm}
\end{figure}
%%%%%%%%%%%%%%%%%%%%%%FIG S1%%%%%%%%%%%%%%%

%%%%%%%%%%%%%%%%%%Table 1%%%%%%%%%%%%%%%%%%%%%
\begin{table*}[t]
\centering
\renewcommand\tabcolsep{8pt}
\caption{Fitting parameters of Eqs\,(\ref{EqS1}) and (\ref{EqS2}) for the thermal conductivity data.}
\label{TableS1}
%\vspace{1mm}
\newlength{\myheight}
\setlength{\myheight}{0.5cm}
\newlength{\myheighta}
\setlength{\myheighta}{1cm}
\begin{tabular}{c||ccccccccc}
\hline
 & $L$ (m)  & $A$ & $D$ (s$^3$) & $B$ (K$^{-1}$s) & $f$  & $g$ (s$^{-1}$) & $h$ & $\frac{\hbar \omega_0}{k_{\rm B}}$ (K) \tabularnewline
\hline
H-Cat  & $0.3\times 10^{-3}$ & $3.64\times 10^{-4}$ & $8.04\times 10^{-41}$ & $2.32\times 10^{-14}$ & $2.65$ & $1.47\times 10^{9}$ & $0.975$ & $8$\tabularnewline
\hline
D-Cat  & $0.3\times 10^{-3}$ & $3.64\times 10^{-4}$ & $8.04\times 10^{-41}$ & $2.32\times 10^{-14}$ & $2.65$ & -- & -- & -- \tabularnewline
\hline
\end{tabular}
\end{table*}
%%%%%%%%%%%%%%%%%%Table 1%%%%%%%%%%%%%%%%%%%%%

%%%%%%%%%%%%%%%%%%%%%%%%%%%%%%%%%%%%%%%%%%%%%%%%%%%%%%%%%%%%%%%%%%
%%%%%%%%%%%%%%%%%%%%%%%%%%%%%%%%%%%%%%%%%%%%%%%%%%%%%%%%%%%%%%%%%%
\section{Specular scattering at sample surfaces}
%%%%%%%%%%%%%%%%%%%%%%%%%%%%%%%%%%%%%%%%%%%%%%%%%%%%%%%%%%%%%%%%%%
%%%%%%%%%%%%%%%%%%%%%%%%%%%%%%%%%%%%%%%%%%%%%%%%%%%%%%%%%%%%%%%%%%

Here, we explain the observed $T^2$-dependence of $\kappa_{\rm ph}$ in H-Cat and D-Cat below $\sim$ 0.25 K. According to the kinetic equation, $\kappa_{\rm ph}$ depends on the heat capacity of the phonons $C_{\rm ph}$, the sound velocity $v_{\rm ph}$ and the mean free path of phonon $l_{\rm ph}$: %as:
\begin{align}
\kappa_{\rm ph} = \frac{1}{3} C_{\rm ph} v_{\rm ph} l_{\rm ph}. \tag{S3}
\label{EqS3}
\end{align}
In general, $l_{\rm ph}$ and $v_{\rm ph}$  become essentially constant with temperature and $C_{\rm ph}$ varies as $C_{\rm ph} \propto T^3$ at temperatures significantly lower than $\mathit{\Theta}_{\rm D}$, leading to the relation $\kappa_{\rm ph} \propto T^3$. However, we found that $\kappa_{\rm ph} \propto T^2$ holds true at sufficiently low temperatures in both H-Cat and D-Cat even though the specific-heat measurements supports $C_{\rm ph} \propto T^3$. %\cite{S.Yamashita2016}. 
Such unexpected behaviours are often observed in glassy materials \cite{ZellerPRB1971} or in high-quality crystals with specular surfaces \cite{HurstPhysRev1969, PohlPRB1982, SutherlamdPRB2003}; in these cases, $l_{\rm ph}$ continues to increase with decreasing temperature. In glassy materials, impurity scattering reduces $l_{\rm ph}$ to below the width or thickness of the sample, whereas in high-quality crystals, $l_{\rm ph}$ can exceed the sample size. Here, we estimated $l_{\rm ph}$ of D-Cat, for which $\kappa = \kappa_{\rm ph}$. As shown in Fig.\,S2, $l_{\rm ph}$ of D-Cat continues to increase with decreasing temperature and extends well beyond the sample boundaries assuming that the value of $v_{\rm {ph}}$ ($\sim$ 1.7 km/s) remains independent of temperature. This result supports the occurrence of specular surface scattering in the present system, which is the cause of the $T^2$-dependence of $\kappa_{\rm ph}$.

%%%%%%%%%%%%%%%%%%%%%%FIG S2%%%%%%%%%%%%%%%
\begin{figure}[b]
\begin{center}
\includegraphics[width=0.9\linewidth]{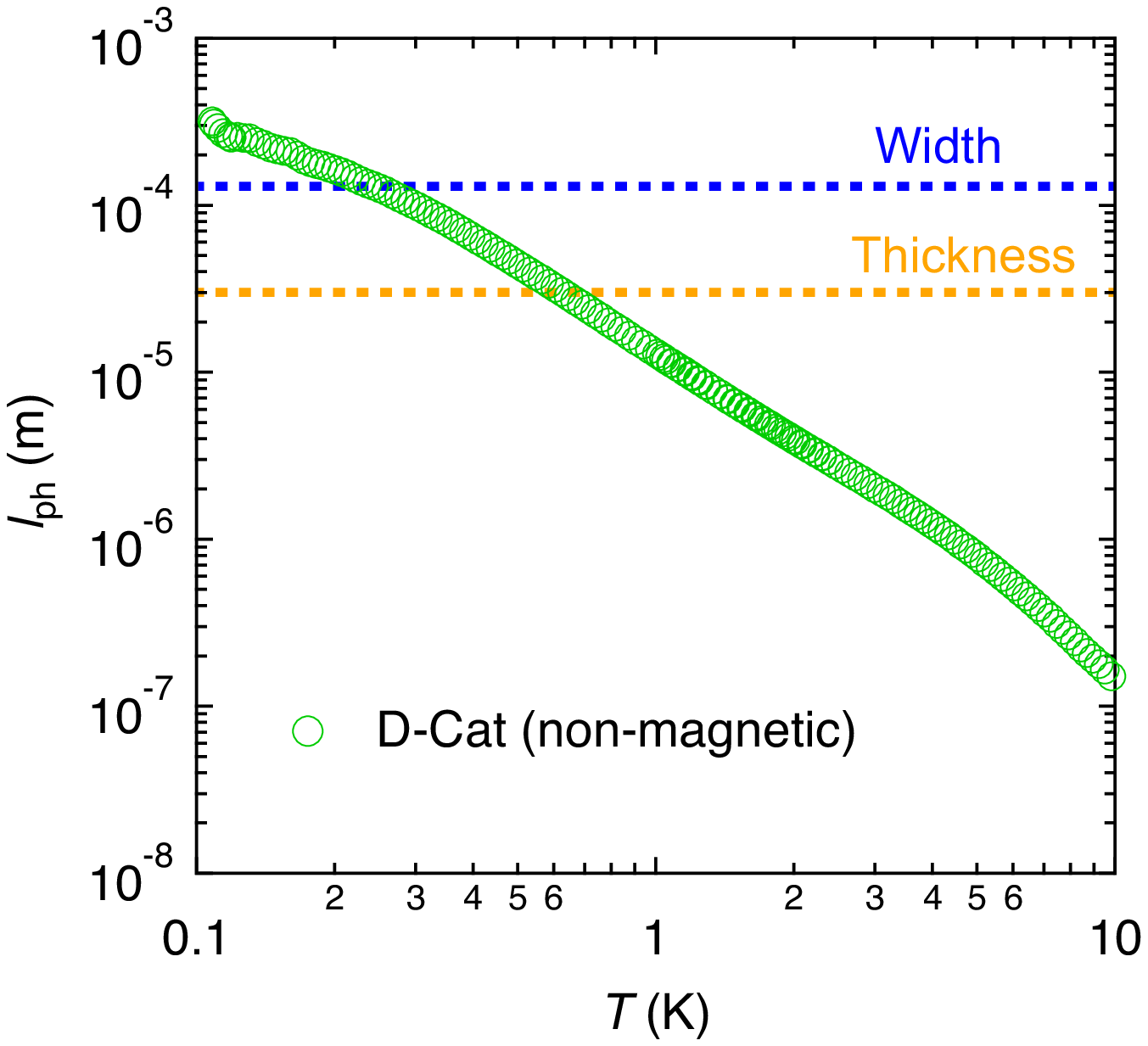}
\end{center}
\vspace{-5mm}
\caption{Temperature dependence of $l_{\rm ph}$ for D-Cat. Blue and orange dotted lines represent the typical width and thickness of single crystals of D-Cat, %investigated here, 
respectively. 
} \label{FigS2}
\end{figure}
%%%%%%%%%%%%%%%%%%%%%%FIG S2%%%%%%%%%%%%%%%

%%%%%%%%%%%%%%%%%%%%%%%%%%%%%%%%%%%%%%%%%%%%%%%%%%%%%%%%%%%%%%%%%%
%%%%%%%%%%%%%%%%%%%%%%%%%%%%%%%%%%%%%%%%%%%%%%%%%%%%%%%%%%%%%%%%%%
\section{Thermal properties of itinerant spin excitations}
%%%%%%%%%%%%%%%%%%%%%%%%%%%%%%%%%%%%%%%%%%%%%%%%%%%%%%%%%%%%%%%%%%
%%%%%%%%%%%%%%%%%%%%%%%%%%%%%%%%%%%%%%%%%%%%%%%%%%%%%%%%%%%%%%%%%%

We investigated the low-lying spin excitation spectrum that characterizes the quantum spin liquid (QSL) states. The spin contribution $\kappa_{\rm sp}$ in H-Cat is extracted by subtracting the phonon contribution $\kappa_{\rm ph}$ from the total thermal conductivity $\kappa = \kappa_{\rm sp} +\kappa_{\rm ph}$. As shown by dashed lines in Fig.\,3b, $\kappa/T$ at low temperatures is well fitted to the formula $\kappa/T = \eta_0 +\beta T$, where $\eta_0$ and $\beta$ are constant coefficients. A finite residual $\eta_0$ of $\sim$ 0.06 W/K$^2$m is clearly resolved in the zero-temperature limit (Fig.\,3b). We note that the residual term is also finite ($\sim$ 0.11 W/K$^2$m; see Fig.\,S3) despite application of the conventional form $\kappa/T = \eta_0 +\beta T^2$. The residual  $\kappa/T$ shows that the spin excitation from the ground state is gapless.

We next estimated the mean free path of the spin excitations $l_{\rm sp}$ \cite{M.YamashitaScience2010}. Analogous to the case of phonons, we assumed that $\frac{\kappa_{\rm sp}}{T} = \frac{1}{3}\frac{C_{\rm sp}}{T}v_{\rm sp}l_{\rm sp}$, where $C_{\rm sp}$ is the specific heat and $v_{\rm sp}$ is the velocity of the spin excitations. Here, we adopted the standard formula of two-dimensional fermion $\frac{C_{\rm sp}}{T} = \frac{\pi^2}{3} k_{\rm B}^2 N_{\rm F} \frac{1}{d}$, where $N_{\rm F}$ is the density of states at the Fermi surface and $d=1.47$ $\mbox{\AA}$ is the interlayer spacing. We also assumed that the linear term $\eta_0$ arises from fermionic excitations and described the energy dispersion as $\epsilon(k)=\hbar v_{\rm sp}k$, where $k$ is the wave number. In this case, the Fermi energy becomes comparable to the spin interaction $J$ and the Fermi wave number is written as $k_{\rm F}=1/a$, where $a$ = 8.36 $\mbox{\AA}$ is the nearest-neighbour spin distance. From the above equations, we obtained $\eta_0=\frac{\pi}{9}\frac{k_{\rm B}^2}{\hbar}\frac{l_{\rm sp}}{ad}$. Because the estimated residual $\eta_0$ is at least 0.06 W/K$^2$m (see the inset of Fig.\,S3), we found that $l_{\rm sp}$ exceeds 120 nm at zero temperature; that is, the spin excitations are mobile to a distance 100 times %as long as 
the inter-spin distance without being scattered. Such highly mobile spin excitations apparently result from the extremely long spin correlation length.

%%%%%%%%%%%%%%%%%%%%%%FIG S3%%%%%%%%%%%%%%%
\begin{figure}[b]
\begin{center}
\includegraphics[width=0.9\linewidth]{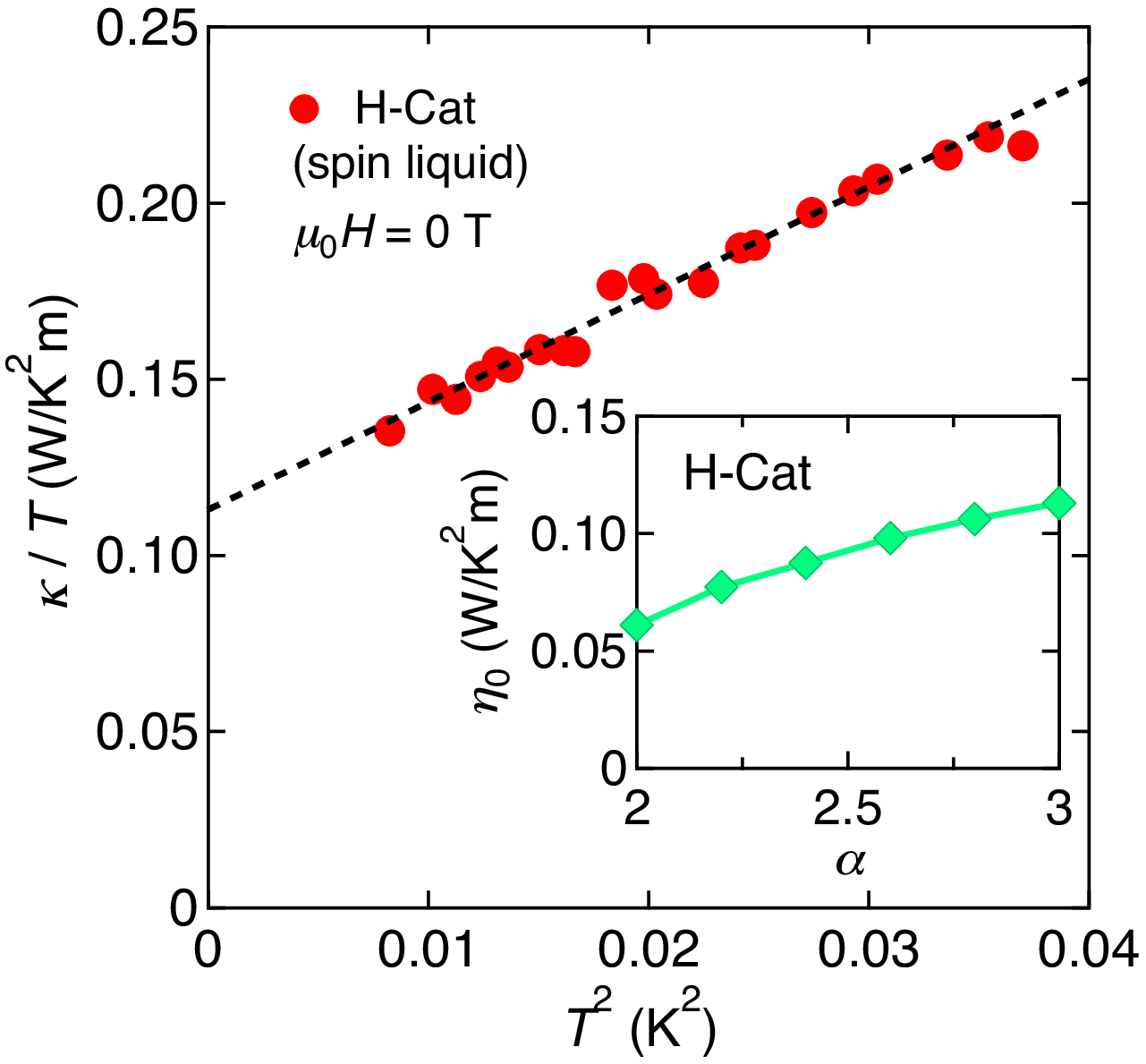}
\end{center}
\vspace{-5mm}
\caption{Low-temperature plot of $\kappa/T$ as a function of $T^2$ in H-Cat. The dashed line represents a fit to $\kappa/T=\eta_0+\beta T^2$, where $\eta_0$ and $\beta$ are constant coefficients. Inset: $\eta_0$ versus the exponent $\alpha$, where $\kappa/T=\eta_0+\beta T^{\alpha -1}$. A finite $\eta_0$ is clearly observed for all $\alpha$. The minimum and maximum values of $\eta_0$ at $\alpha =$ 2--3 are approximately 0.06 and 0.11 W/K$^2$m, respectively.
} \label{FigS3}
\end{figure}
%%%%%%%%%%%%%%%%%%%%%%FIG S3%%%%%%%%%%%%%%%

%%%%%%%%%%%%%%%%%%%%%%%%%%%%%%%%%%%%%%%%%%%%%%%%%%%%%%%%%%%%%%%%%%
%%%%%%%%%%%%%%%%%%%%%%%%%%%%%%%%%%%%%%%%%%%%%%%%%%%%%%%%%%%%%%%%%%
\section{Estimation of distance from Mott transition}
%%%%%%%%%%%%%%%%%%%%%%%%%%%%%%%%%%%%%%%%%%%%%%%%%%%%%%%%%%%%%%%%%%
%%%%%%%%%%%%%%%%%%%%%%%%%%%%%%%%%%%%%%%%%%%%%%%%%%%%%%%%%%%%%%%%%%

The distance from a Mott transition can be estimated from the ratio of the on-site Coulomb repulsion $U$ to the transfer integral $t$, which is given by $U/t\sim t/J$, where $J$ is the spin interaction. The value of $t$ is calculated to be 30--40 meV for H-Cat \cite{TsumurayaPRB2015}, $\sim$ 50 meV for $\kappa$-(BEDT-TTF)$_2$Cu$_2$(CN)$_3$ \cite{KandpalPRL2009, NakamuraJPSJ2009} and 25--40 meV for EtMe$_3$Sb[Pd(dmit)$_2$]$_2$ \cite{TsumurayaJPSJ2013}. Note that $t$ is comparable among the three materials. By contrast, $J/k_{\rm B}$ of H-Cat is estimated as $\sim$ 80 K \cite{IsonoPRL2014} and as $\sim$ 1/3 in the other two compounds ($J/k_{\rm B} \sim$ 250 K for both $\kappa$-(BEDT-TTF)$_2$Cu$_2$(CN)$_3$ \cite{ShimizuPRL2003} and EtMe$_3$Sb[Pd(dmit)$_2$]$_2$ \cite{ItouPRB2008}). Consequently, H-Cat has a relatively large $U/t$; this indicates a large distance from the Mott transition.

%%%%%%%%%%%%%%%%%%%%%%%%%%%%%%%%%%%%%%%%%%%%%%%%%%%%%%%%%%%%%%%%%%
%%%%%%%%%%%%%%%%%%%%%%%%%%%%%%%%%%%%%%%%%%%%%%%%%%%%%%%%%%%%%%%%%%
\section{Theoretical model of an electron-proton coupled system}
%%%%%%%%%%%%%%%%%%%%%%%%%%%%%%%%%%%%%%%%%%%%%%%%%%%%%%%%%%%%%%%%%%
%%%%%%%%%%%%%%%%%%%%%%%%%%%%%%%%%%%%%%%%%%%%%%%%%%%%%%%%%%%%%%%%%%

We introduce a minimal model that describes a coupling between the $\pi$-electron and proton degrees of freedom in H-Cat and D-Cat. The model Hamiltonian is given by 
\begin{align}
{\cal H} = {\cal H}_{\rm e} + {\cal H}_{\rm p}, \tag{S4}
\label{eq:hamil}
\end{align}
where the first term ${\cal H}_{\rm e}$ describes the $\pi$-electron system and the second term ${\cal H}_{\rm p}$ represents the proton system coupled with the $\pi$-electrons. The first term is given by the extended Hubbard model (EHM): 
%The extended Hubbard model (EHM) was adopted in the first term ${\cal H}_{\rm e}$ as 
\begin{align}
{\cal H}_{\rm e} 
&= t_{\rm d} \sum_{i \sigma}(c^{\dagger}_{i \alpha \sigma}c_{i \beta \sigma} + {\rm H. c.})
+ \sum_{\langle ij \rangle \sigma} t^{\mu\mu'}_{ij}(c^{\dagger}_{i \mu \sigma}c_{j \mu' \sigma} + {\rm H. c.}) \notag \\
&+ U_0\sum_{i \mu}n_{i \mu \uparrow}n_{i \mu \uparrow}
+ V_{\rm d} \sum_{i}n_{i \alpha}n_{i \beta}
+ \sum_{\langle ij \rangle}V_{ij}^{\mu\mu'}n_{i\mu}n_{j\mu'}, \tag{S5}
\label{eq:hamil_e}
\end{align}
where $c_{i\mu\sigma}$ is the annihilation operator of a hole with spin $\sigma=(\uparrow, \downarrow)$ at the $\mu$ $(= \alpha, \beta)$ molecule of the $i$-th dimer, $n_{i\mu} \equiv \sum_{\sigma} n_{i\mu\sigma} \equiv \sum_{\sigma} c^{\dagger}_{i\mu\sigma}c_{i\mu\sigma}$ is the number operator and $t_{\rm d}$ ($V_{\rm d}$) is the intra-dimer transfer integral (Coulomb interaction) between the $\alpha$ and $\beta$ molecules inside a dimer. Moreover, $t^{\mu\mu'}_{ij}$ ($V_{ij}^{\mu\mu'}$) is the inter-dimer transfer integral (Coulomb interaction) between the $\mu$ molecule of the $i$-th dimer and the $\mu'$ molecule of the $j$-th dimer and $U_0$ is the intra-molecular Coulomb interaction (see Fig.\,S4a). %The first and second terms in Eq.\,(\ref{eq:hamil_e}) represent the intra- and inter-dimer electron hoppings, respectively, and the third and fourth terms describe the electron-electron interactions on a molecule and between the two molecules inside a dimer, respectively.
The second term ${\cal H}_{\rm p}$ in Eq.~(\ref{eq:hamil}) is described by
\begin{align}
{\cal H}_{\rm p} = 2 t_{\rm p}\sum_{i} P_i^{x} + \frac{1}{2}g \sum_{\langle ij \rangle} (n_{i \beta} - n_{j \alpha})P_{i}^{z}, \tag{S6}
\end{align}
where $t_{\rm p}$ is the proton tunneling amplitude (proton-fluctuation strength) \cite{Blinc60}, $g$ ($>0$) is the coupling constant between the hole and the hydrogen atom and $\bm P_i$ is a pseudo-spin operator with an amplitude of $1/2$ describing the proton degree of freedom at the $i$-th hydrogen bond. Here, we define %that 
the eigenstates of $P^{z}$ as $\left| + \right\rangle$ and $\left| - \right\rangle$ , respectively. In these states, the hydrogen atom is localized at the right and left sides of the double potential minima, respectively (see Fig.\,S4b). The eigenstates of $P^{x}$, denoted by $\left| A \right\rangle = (\left| + \right\rangle + \left| - \right\rangle)/\sqrt{2}$ and $\left| B \right\rangle = (\left| + \right\rangle - \left| - \right\rangle)/\sqrt{2}$, represent the bonding and anti-bonding states, respectively (see Fig.\,S4b). This pseudo-spin representation has been applied to hydrogen-bonded systems such as the ferroelectric KH$_2$PO$_4$ (KDP) \cite{Vaks66,Larkin69}, where displacive deformation of PO$_4$ tetrahedrons is coupled with O-H-O hydrogen dynamics.

%%%%   OLD version %%%%
%Here, we introduce a pseudo-spin operator $\bm P_i$ with an amplitude of $1/2$ \cite{Vaks66,Larkin69} to describe the proton degree of freedom at the $i$-th hydrogen bond (see Figs\,S4a and b). We define that the eigenstates of $P^{z}$, denoted by $\left| + \right\rangle$ and $\left| - \right\rangle$, represent the states where the hydrogen atom is localized at the right and left sides of the double potential minima, respectively, while the eigenstates of $P^{x}$, denoted by $\left| A \right\rangle = (\left| + \right\rangle + \left| - \right\rangle)/\sqrt{2}$ and $\left| B \right\rangle = (\left| + \right\rangle - \left| - \right\rangle)/\sqrt{2}$, represent the bonding and anti-bonding states, respectively. The second term ${\cal H}_{\rm p}$ in Eq.~(\ref{eq:hamil}) is defined by 
%\begin{align}
%{\cal H}_{\rm p} = 2 t_{\rm p}\sum_{i} P_i^{x} + \frac{1}{2}g \sum_{\langle ij \rangle} (n_{i \alpha} - n_{j \beta})P_{i}^{z}, \tag{S6}
%\end{align}
%where $t_{\rm p}$ is the proton tunneling amplitude (proton-fluctuation strength) \cite{Blinc60} and $g$ is the coupling constant between the hole and the hydrogen atom.

%%%%%%%%%%%%%%%%%%%%%%FIG S4%%%%%%%%%%%%%%%
\begin{figure}[t]
\begin{center}
\includegraphics[width=\linewidth]{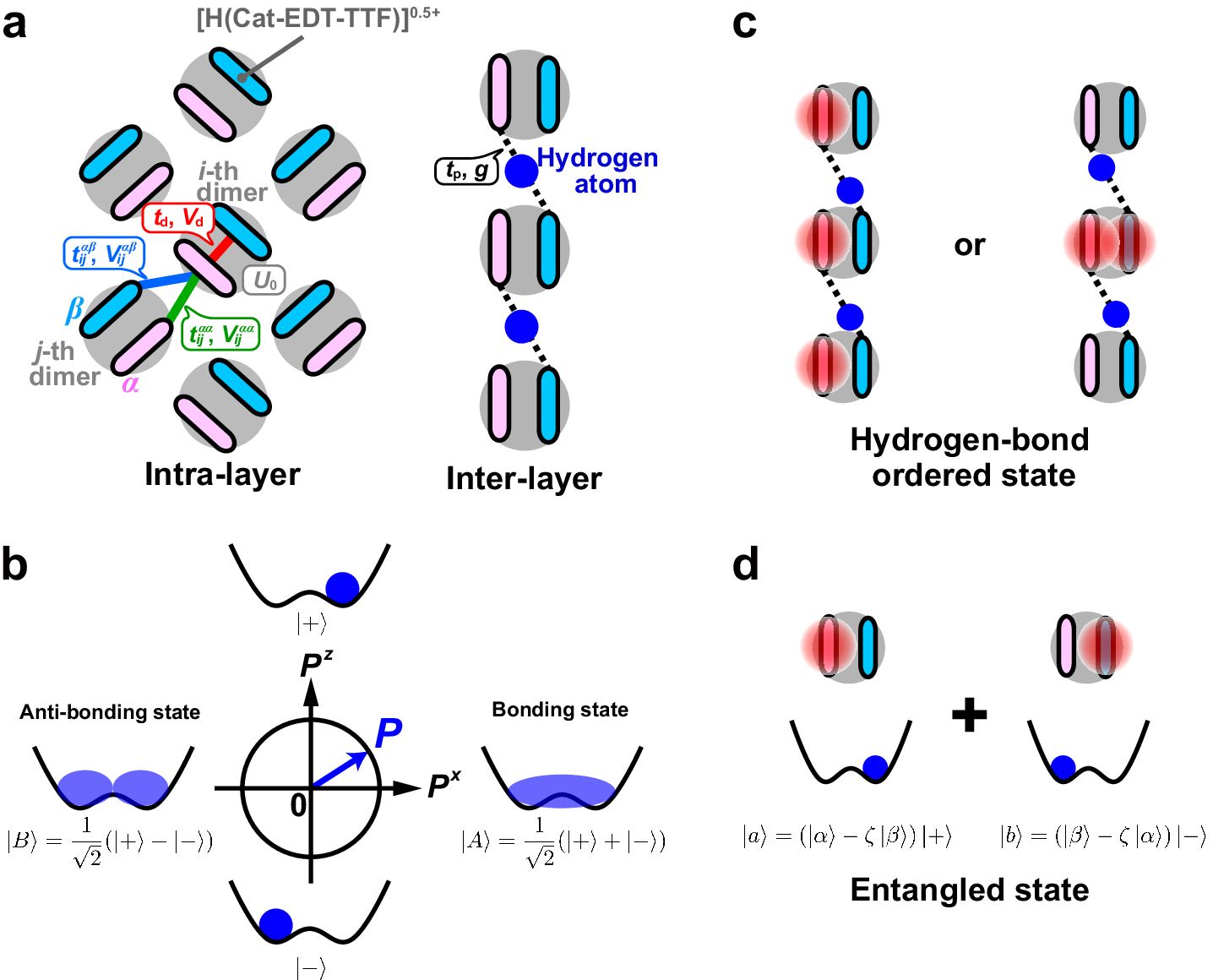}
\end{center}
\vspace{-3mm}
\caption{{\bf a,} Schematic lattice structures of an intra-layer (left) and inter-layer (right) for H-Cat.
% viewed within the $b$-$c$ plane (left) and along the $a^{\ast}$ axis (right).
The ellipses represent the [H(Cat-EDT-TTF)]$^{0.5+}$ molecules and the two molecules in a dimer (gray circle) are labelled $\alpha$ and $\beta$. The red, green and blue lines represent the transfer integrals and the Coulomb interactions introduced in the EHM (see text). The black dotted lines indicate proton tunneling and the coupling constant between the hole and the hydrogen atom. 
{\bf b,} Pseudo-spin directions in the $P^{x}$--$P^{z}$ plane and the corresponding states of the hydrogen atom.
{\bf c, d,} Schematic views of hydrogen-bond ordered states ({\bf c}) and an entangled state ({\bf d}) expected in the present theoretical model. For simplicity, we consider the case of $\zeta=0$. The size of red circles represents a relative hole density.}
\label{FigS4}
\vspace{-3mm}
\end{figure}
%%%%%%%%%%%%%%%%%%%%%%FIG S4%%%%%%%%%%%%%%%

We now consider a 1D chain of $\pi$-dimers and hydrogen bonds in the strong electron correlation regime, %$U_0/t \gg 1$, 
where the on-site Coulomb interaction is larger than the inter-dimer Coulomb interaction. In this situation, for relatively small $t_{\rm p}$, hydrogen-bond order occurs concomitantly with charge order inside of the $\pi$-dimer, leading to two types of ordering patterns. In one case, the hydrogen atoms are located at the same oxygen side for two O-H-O hydrogen bonds and each hole is distributed to the same molecular side of each dimer (see the left panel of Fig.\,S4c). In the other case, each hydrogen atom is located at the opposite oxygen side, leading to a state where one dimer is occupied by two holes and the other is empty (see the right panel of Fig.\,S4c); this state corresponds to the non-magnetic D-Cat.

By contrast, for relatively large $t_{\rm p}$, any hydrogen order does not occur owing to the strong proton tunneling effect, which corresponds to the state for H-Cat. To investigate this state more closely, we consider the simplest case, in which an isolated dimer occupied by one hole is coupled to a proton. In this case, the ground-state wave function is given by an entangled state (see Fig.\,S4d)
$\mathit{\Phi}_{1,2}=\left|a\right\rangle+\left| b\right\rangle = (\left| \alpha \right\rangle -\zeta \left| \beta \right\rangle)\left| + \right\rangle + (\left| \beta \right\rangle -\zeta \left| \alpha \right\rangle) \left| - \right\rangle$. Here, $\left| \alpha \right\rangle$ and $\left| \beta \right\rangle$ represent the states where the $\pi$ electron is localized at the $\alpha$ and $\beta$ molecules inside the dimer, respectively. In addition, $|\zeta|$ ($< 1$) expresses the degree of hybridization of $\left|\alpha\right\rangle $ and $\left|\beta\right\rangle $ induced by $t_{\rm d}$. We stress that when $t_{\rm p}\neq 0$, $\mathit{\Phi}$ cannot be described by a direct product $\mathit{\Phi}_{\rm{h}} \otimes \mathit{\Phi}_{\rm{p}}$, where $\mathit{\Phi}_{\rm{h(p)}}$ is the hole (proton) wavefunction expressed as a linear combination of $|\alpha\rangle$ and $|\beta\rangle$ ($|+\rangle$ and $|-\rangle$), whereas at $t_{\rm p}=0$, the eigenstates of $\left| a \right\rangle$ and $\left| b \right\rangle$ are degenerate. Such a proton--electron entanglement in the presence of $t_{\rm p}$ has been  confirmed in numerical calculations of the exact diagonalization method in cluster systems. Thus, proton tunneling $t_{\rm p}$ is an important factor for determining the ground states in the present system. To clarify the origin of the QSL state and the effects of the underlying proton--electron entanglement, further theoretical studies are required.

%%%%%%%%%%%References